\documentstyle[12pt,aaspp4]{article}
\def\ea{{\it et al.} }
\def\ref{\par\noindent\hangindent 20 pt}

\def\aro{$\alpha_{RO} \; $}
\def\sec{^{\prime\prime\!\!}}
\def\cosmo{H$_0=50$ km/s/Mpc and q$_0=0\;$ }

\newcommand{\simlt}{\raisebox{-3.8pt}{$\;\stackrel{\textstyle < }{\sim }\;$}}
\def\sec{^{\prime\prime\!\!}}

\begin{document} 

\input{psfig.tex}

\title{\bf HST Observations of the Optical Jets of PKS 0521-365, 3C371, 
and PKS 2201+044}

\author{Riccardo Scarpa}
\authoraddr{Space Telescope Science Institute, 3700 San Martin Dr., 
Baltimore, MD 21218, USA}
\authoremail{scarpa@stsci.edu}
\affil{Space Telescope Science Institute}

\author{C. Megan Urry} 
\authoraddr{Space Telescope Science Institute, 3700 San Martin Dr., Baltimore,
MD 21218, USA}
\authoremail{cmu@stsci.edu}
\affil{Space Telescope Science Institute}

\author{Renato Falomo} 
\authoraddr{Osservatorio Astronomico di Padova, Vicolo dell'Osservatorio 5, 
35122 Padova, Italy}
\authoremail{falomo@astrpd.pd.astro.it}
\affil{Astronomical Observatory of Padova}

\author{Aldo Treves} 
\authoraddr{University of Milan at Como, Via Lucini 3, Como Italy}
\authoremail{treves@uni.mi.astro.it}
\affil{University of Insubria, Como}

\begin{abstract}

HST observations have led to the discovery of the optical counterpart
of the radio jet of PKS 2201+044, and to a detailed analysis of the
optical jets of PKS 0521-365 and 3C371.
At HST spatial resolution these jets are well resolved, displaying knotty
morphologies. When compared with radio maps of appropriate resolution, a
clear one-to-one correspondence between optical and radio structures is
found, showing that all detected optical structures are indeed related to
the radio synchrotron emission. 
Photometry of the brightest knots shows that the
radio-to-optical spectral index and the derived intensity of the
equipartition magnetic field are approximately constant along the
jet. Thus, present observations suggest that the electron energy
distribution does not change significantly all along the jet.

{\it Subject headings}: galaxies: jets -- galaxies: individual (PKS 0521-365,
3C 371, PKS 2201+044) -- BL Lacertae objects

\end{abstract}

\section{Introduction.}

Radio-emitting synchrotron jets are commonly observed in radio
galaxies and quasars.  In contrast, the optical counterparts of these
structures are observed only rarely (e.g., O'Dea \ea 1999; Scarpa \&
Urry 1999 and references therein). 
The high spatial resolution of the {\it Hubble Space
Telescope} (HST) allows the study of known optical jets with detail
comparable to VLA radio maps, and has led to the discovery of new
optical jets.  HST is particularly useful for finding short
jets that develop within few arcseconds from the luminous nucleus and/or
are fully contained within a luminous host galaxy.  The jet emission
is typically confined to a narrow area, so at higher spatial
resolution its surface brightness is higher.  This facilitates
detection closer to the nucleus, increasing the probability of
detecting either shorter jets or jets more closely aligned with the
line of sight. Not surprisingly, the number of known jets has almost
doubled since the introduction of HST (e.g., Sparks \ea 1995; Ridgway
\& Stockton 1997).

Detection of optical emission and high resolution optical imaging of
radio jets put important constraints on jet physics. 
For example, electrons radiating at $\nu \sim 10^{15}$ Hz in a 
typical equipartition magnetic field of $10^{-4}$~G 
have a radiative cooling time of only $t_{1/2}\sim 1000$ years,
yet optical jets are tens of thousands of light years long.
(e.g., PKS 0521-365, Danziger \ea 1979). 

Here we present new HST images
\footnote{Based on observations made with the NASA/ESA
Hubble Space Telescope, obtained at the Space Telescope Science
Institute, which is operated by the Association of Universities for
Research in Astronomy, Inc., under NASA contract NAS~5-26555.}
of three relativistic jets which probe the shortest-lived radiating
particles.  The observations were obtained as part of our HST
snapshot survey of BL Lac objects (Urry et al.  1999, Scarpa et
al. 1999) and so were shorter than the ideal for studying synchrotron
jets; however, the large number of low redshift BL Lac observed greatly
increased our chance of detecting new jets. Indeed, in PKS 2201+044, the
104$^{th}$ of 109 objects observed, we discovered a new
optical jet. We also imaged two well-known BL Lacertae objects with
previously known optical jets, PKS 0521-365 and 3C371, which are
relatively nearby and can be studied in great detail with HST. 

Throughout the paper \cosmo were used.

\section{Observations and Data Analysis}

Observations were carried out with the Hubble Space Telescope in
snapshot mode, using the Wide Field and
Planetary Camera 2 (WFPC2) through filters F702W (for PKS 0521-365 and
PKS 2201+044) and F555W (for 3C371). 
Targets were centered on the PC chip, which has pixels
$0.046$ arcsec wide.  To obtain a final image well exposed both in
the inner, bright nucleus and in the outer fainter regions, we set up
a series of exposures with duration ranging from a few to hundreds of seconds,
for total integration times of 305, 302, and 610 seconds for
PKS 0521-365, 3C371, and PKS 2201+044, respectively.

After standard reduction (flat--field, dark and bias
subtraction, and flux calibration carried out as part of the standard
HST pipeline processing), images were combined using the task ``CRREJ''
available within IRAF.  This task stacks all images while removing
cosmic ray events.  Final images were flux calibrated following the
prescription of Holtzman et al. (1995, their Eq.~9 and Table~10).

Although readily visible in the direct HST image, to be fully studied
the jet must be separated from the contaminating light of the host
galaxy and the central, luminous point source.  We therefore first
subtracted a suitably scaled point spread function (PSF) 
model computed using Tiny Tim
software (Krist 1995), plus an estimate of large-angle 
scattered light (Scarpa \ea 1999). 
With the IRAF task ``ELLIPSE'' we then fitted the residual 
light with a two-dimensional elliptical galaxy model,
convolved with the PSF.  Extraneous features, including the jet
itself, were masked before fitting the isophotes. 
(For details on host galaxies
see Scarpa \ea 1999 and Falomo \ea 1999.) The final jet image
was obtained after subtracting both PSF and galaxy models.

\section{Description of the Three Jets}

\subsection{PKS 0521-365} 
This well-studied nearby object (z=0.0554; Danziger \ea
1979) is one of the most remarkable extragalactic objects known, as it
shows a variety of nuclear and extra-nuclear phenomena. First
classified as an N galaxy, then as a BL Lac, now it
shows the strong narrow and broad emission lines typical of Seyfert 1
galaxies (Ulrich 1981; Scarpa, Falomo \& Pian 1995). Most interesting,
the source has a prominent radio and optical jet (Danziger \ea 1979; Keel 1986;
Macchetto \ea 1991; Falomo 1994), which resembles that of the nearby
radio galaxy M87 (i.e., Sparks, Biretta \& Macchetto 1994).

The optical jet as observed at HST resolution (Fig. 1) has a structure
characterized by a bright feature near the nucleus (marked with an
``F''), which is however too close to the bright central point source
to be readily studied. It corresponds to the base of the jet visible
in the UV image obtained with the HST Faint Object Camera (Macchetto
\ea 1991). Farther out, a prominent bright spot (called ``A''
following Falomo 1994), a series of smaller knots and a final
bright spot (``C'') are also visible.  Structure ``B'' is resolved
into at least three distinct knots aligned along the jet axis.
Structure ``C'' (also called the ``red tip'') is $9\sec .2$ from the
nucleus and has integrated magnitude $m_R=21.3\pm 0.2$ mag.  It is
clearly resolved even if its symmetric shape makes it appear
point-like, and its radial profile is consistent with an elliptical
galaxy. This confirms the finding of Falomo (1994), and together with
the absence of radio emission (Keel 1986) and detectable optical
polarization (Sparks, Miley \& Macchetto 1990), suggests the ``red
tip'' is a background galaxy (or small companion) rather than part of
the jet, in spite of the almost perfect alignment.

Finally, we note that in the optical there is no indication of either
a counter-jet or the 2cm bright hot spot reported by Keel (1986) S-E
of the nucleus, which was also not detected in deeper ground based
observations (Falomo 1994).

We were able to measure the apparent magnitude of 5 bright knots.  In
Table 1 we report the apparent R magnitude, the distance of each
feature from the central point source, and the jet width at these
positions.  The reported jet width is the average full width at zero
intensity estimated averaging on strips 10 pixels wide. Errors on the
jet width are $\pm 0\sec .2$ for all values.  The total magnitude of
the jet, excluding knots ``C'' and ``F'', is m$_R=19.9\pm0.2$ mag.

The jet is resolved also orthogonal to the axis. In feature ``D'' two
knots are clearly visible, ``E'' is tilted with respect to the
jet axis, and around ``B'' several secondary maxima of light are
detected.  Indeed, compared with lower resolution ground-based images,
in its terminal region the jet looks much more diffuse than previously
reported.  At the lowest surface brightness reached in our
observations, the jet has an almost constant width (as is also the
case in the deeper image obtained by Falomo 1994). Even in knot ``A'',
where the jet is very bright, the full width at zero intensity is not
significantly larger than in the other zones. Only at the very end, at
``B1'', is the jet narrower. Present data therefore suggest the plasma
is moving in a well defined cylindrical funnel.

The optical morphology of the jet agrees very 
well with what is observed at 2cm
(Keel 1986).  The general shape and the total length of the radio jet
are almost exactly reproduced in the optical. This is particularly true
for knot ``A'', which is well resolved in both bands (Fig. 2).

The 2 cm flux was given by Keel (1996) for the whole jet, knot ``A'',
and the jet minus ``A''. For other three knots we estimated 2cm flux
directly from the radio map of Keel (1986). We then evaluated the
radio-to-optical spectral index for four knots and the full jet (Table
1). It turns out that \aro is relatively stable along the jet, indeed
all values are consistent within the errors. This is simply a
quantitative confirmation of the general one-to-one morphological 
correspondence between the two bands.

Assuming equipartition between particles and magnetic field, the
internal magnetic field can be estimated knowing the total flux and
volume of a radio source. In our case, the size of the emitting knots can be
directly measured on the image and transformed into volume assuming
symmetry. We report in Table 1 the volume $V$ of the whole jet and of
each knot. The jet is assumed to be a cylinder of diameter 1.5 arcsec
and length 5 arcsec. For the knots we assumed spherical symmetry,
reporting in Table 1 the volume of the sphere which has projected
surface equal to the knot surface.  This was done because the jet is
resolved orthogonal to its axis, with the knots occupying only a
fraction of the jet width.

The complete radio-to-optical flux energy distribution of the
synchrotron emission from the jet alone is not known. We therefore
assume the synchrotron emission from jet and knots is described by
power laws, with spectral indices as reported in Table 1, extending from
$\nu \sim 10^7$ to $10^{15}$ Hz. The upper frequency limit is set by
the HST FOC data at 3200\AA\ (Macchetto \ea 1991) and has basically no
effect on the estimated equipartition magnetic field. 
With these assumptions, the
jet has almost constant $B_{eq} = 6\times10^{-5}$ G along its length (Table
1). Magnetic fields of this intensity are commonly observed in radio
sources, and our values are fully consistent with the value of
$B_{eq}=5\times10^{-5}$ G reported by Keel (1986).

\subsection{3C371} 

This extragalactic radio source has been the subject of intense study
at many frequencies. It was first classified as an optically violently
variable quasar (Angel \& Stockman 1980) and then as a radio-selected
BL Lacertae object (Giommi \ea 1990).  The radio morphology is
dominated by a one-sided jet emanating from the central core and
extending up to 25 arcsec from the nucleus, ending at one of the two
radio lobes (Wrobel \& Lind 1990). This morphology is characteristic
of a Fanaroff-Riley type II radio source.  In the optical band 3C371
is hosted by a large elliptical galaxy (UGC 11130, Condon \& Broderick
1988) at $z=0.0508$ (Sandage 1973), surrounded by a small cluster of
galaxies (Stickel, Fried \& K\"uhr 1993).  Recently the optical
counterpart of the radio jet was discovered by Nilsson \ea (1997).  As
observed from the ground, the optical jet extends only a few arcsec
from the nucleus, ending in a bright knot where the radio jet
apparently changes direction.

At HST resolution, the jet is fully resolved also perpendicular to the
jet axis. At least three bright knots are clearly visible
(Fig. 3). Near the center, the jet starts with a bright structure
(``C''), followed by a much smaller knot (``B'') which is elongated
orthogonal to the jet. Further out is the broadest structure, knot
``A'', $\sim 1\sec .3$ wide.  In our HST image, the optical jet
appears to extend beyond ``A''; the narrow structure named ``D'' in
Fig. 3 coincides with a faint feature visible in the Nilsson \ea
(1997) image. The beginning of feature ``D'' is possibly also visible
in the MERLIN map of Akujor \ea (1994).

Comparing optical and radio emission a general one-to-one
matching of all knots is found.
The observed radio jet (Wrobel \& Lind 1990) is several times longer
than the optical jet. Compared to PKS 0521-365 and M87, this is the
most obvious difference. It must be noted that the radio flux fades
significantly after knot ``A''. Wrobel \& Lind (1990) report a peak
flux of 3.8 mJy from the final hot spot of the radio jet.  Assuming
\aro $= -0.7$ (Table 1) the expected flux in V-band is $\simlt
5.7\times 10^{-4}$ mJy, roughly 1 magnitude fainter than the limit of
our observation. Hence, present data are consistent with a constant
$\alpha_{RO}$ and general correspondence of radio and optical
emission, and deeper images may well be able to detect optical
emission up to the very end of the radio jet. On the contrary, if the
optical jet actually ends at ``A'', it would be even more interesting,
offering an important benchmark for theoretical models to explain the
propagation of electrons and magnetic field through the interstellar
medium. Deeper optical observations are therefore highly desirable.

Under the same set of assumptions adopted for PKS 0521-365 about
synchrotron power law spectrum and emission volume, we estimate the
physical properties of the jet.  In Table 1 the V magnitude and volume
for the whole jet and main knots are reported. Knot ``C'' is too close
to the central point source so we do not attempt to measure its
luminosity. To estimate the volume of the jet we modeled it as a
cylinder with diameter of 1 and length $3.5$ arcsec, while spherical
symmetry was assumed in computing the volume of knots.  The radio flux
at 1.6 GHz (Akujor \ea 1994) is also quoted.

The same value of the radio-to-optical spectral index is found both
for the whole jet and knot ``A''.  Considering the H- and B-band data
from Nilsson \ea (1997) together with our V-band data, the
near-IR-optical spectral index of knot ``A'' is
$\alpha_{IR-O}=-0.63\pm 0.2$. This is fully consistent with \aro,
indicating no steepening of the radiation spectrum at high frequencies.

The equipartition magnetic field is basically constant along the jet
and has average intensity $<B_{eq}>=6.5\times10^{-5}$~G, similar to what was
found for PKS 0521-365.

\subsection{PKS 2201+044}

This source is a nearby ($z=0.028$) BL Lac object (Angel \& Stockman
1980) hosted in a prominent elliptical galaxy (Falomo 1996; Scarpa \ea
1999).  The optical spectrum shows emission and absorption lines and
is similar to Seyfert 1 spectra (Veron-Cetty \& Veron 1993; Falomo,
Scarpa \& Bersanelli 1994). The associated radio source has a total
extent of 5.6 arcmin (Ulvestad \& Johnson 1984), and is core dominated
(Preston \ea 1985).  The radio properties of PKS 2201+044 have been
throughly investigated by Laurent-Muehleisen \ea (1993), who found a
core-jet morphology, with the jet extending for more than 10 kpc. A
bright radio hot spot was also found $1.5$ kpc (2 arcsec) from the
nucleus.

At HST resolution the optical source is fully resolved into nucleus,
host galaxy, and jet. Thanks to the superior spatial resolution, some
structures of the jet are directly 
visible above the strong background of the host
galaxy. In Figure 4 we show the residuals after subtraction of the best
fit galaxy model. The jet structure appears at position angle
$310^\circ$, with the most prominent knot $2\sec .12$ (1.57 kpc) from
the nucleus. Comparing our data to the 5GHz radio map of
Laurent-Muehleisen \ea (1993), a clear correspondence of this bright
optical knot with the brightest radio knot is found, strongly
supporting the association of the optical structure with the
synchrotron radio jet (Fig. 5).  Unfortunately, the resolution of the
radio map is insufficient to resolve further structures.

The brightest optical knot is clearly resolved both along and perpendicular
to the jet axis, and has quite a
complicated structure, extending over a region of $\sim0.5\times 0.9$
arcsec$^2$. Between the knot and the nucleus some weaker knots may also be
present. 

The integrated magnitude of knot ``A'' is $m_R=24.2\pm 0.1$~mag ($5.7\times
10^{-4}$ mJy). Its radio flux is $\sim 9$ mJy at 5GHz, corresponding
to radio-to-optical spectral index $\alpha_{RO}=-0.85$, fully
consistent with the values found in others optical jets (Crane \ea
1993; Scarpa \& Urry 1999).

Assuming the knot is a sphere of radius 0.4 arcsec, and adopting the
same set of hypotheses as for the previous two jets, the equipartition
magnetic field is $B_{eq}=8\times 10^{-5}$~G, again similar to what is
found in others jets.

\section{Conclusions}

We have described HST observations of three synchrotron jets detected
at optical wavelengths in three BL Lac objects. In two cases, PKS 0521-365 
and 3C371, the jet is fully resolved and several knots can be studied
separately. In spite of the clumpiness of the emission, both jets are
rather homogeneous, with basically constant values of \aro $\sim -0.7$
and equipartition magnetic field along the length of the jet.  For
knot ``A'' in 3C371, a similar spectral index is also found at
near-IR-optical frequencies, indicating no significant aging of the
radiating electrons.

We also reported the discovery of an optical counterpart of the radio
jet of PKS 2201+044. The brightest knot, at projected distance $1.57$
kpc from the nucleus, is fully resolved in the optical. It corresponds
precisely to the brightest radio knot, confirming its association with
the synchrotron jet. Estimates of $\alpha_{RO}=-0.85$ and
equipartition magnetic field $B_{eq}=8\times 10^{-5}$ G are similar to
values generally found in other jets. However, to fully understand the
morphology of this jet, deeper optical images and higher resolution
radio maps are needed.

We note that all three sources exhibit strong emission lines in
their optical spectrum, and in PKS 2201+044 and PKS 0521-365 broad
emission lines are also observed (Ulrich 1981; Veron-Cetty \& Veron 1993;
Scarpa, Falomo \& Pian 1995). This may suggest that in these
objects the beaming is modest, and in fact there are indications of
this in the case of PKS 0521-365 (Pian \ea 1996). A modest
beaming may be consistent with an easier observability of the optical
jet. However our sample is obviously too restricted to elaborate further.

At present, including the newly discovered jet in PKS 2201+044, a
total of 14 optical jets are known. This number is not big, but it
starts to be large enough to enable decent statistical
consideration. It is well known that in all these jets the inferred
electron lifetimes are much smaller than needed to explain the jet
length, so that reacceleration along the jet seems unavoidable. This
is also true for the jet in PKS 2201+044. However,
the observed constancy of $\alpha_{RO}$ is difficult to explain if
electrons are reaccelerated by first-order Fermi mechanism at
strong shocks (Heinz \& Begelman 1997, and references 
therein), so that present
observations add difficulties to the already hard problem of finding a
plausible physical mechanism for electron reacceleration.
The possibility of electrons being transported in a loss-free  channel
inside the jet (Owen, Hardee \& Cornwell 1989) is difficul to accept, 
because  HST observation of M87 have shown the optical
emission is centered near the axis of the jet rather than on the edge
(Spark, Biretta \& Macchetto 1996), which appear to be the case also 
for PKS 0521-365.
It is possible that the presence of relativistic motion on kiloparsec
scales is common, as supported by direct
observations with HST of superluminal proper motion in M87 (Biretta
\ea 1999). If this is generally the case, then relativistic beaming of
the emission is unavoidable, implying that both the intrinsic
luminosity of the jet and the equipartition magnetic field are smaller
than what we estimated, while electron lifetimes are longer and the
necessity for reacceleration is weakened. 

An alternative hypothesis
is to renounce the basic assumption of equipartition of
energy between relativistic particles and magnetic field; the electron
lifetimes are longer by approximately the factor $(B_{eq}/B)^{1.5}$.
Using a statistical approach we address general
considerations of jets power, beaming, and physical conditions in a
companion paper (Scarpa \& Urry 1999).

\acknowledgements

It is a pleasure to thanks G. Ghisellini and F. Macchetto for helpful
and encouraging comments. We also thanks the 
referee W. C. Keel for helpfull comments
and for providing its radio map of PKS 0521-365 in fits format.
Support for this work was provided by NASA through grant 
GO-06363.01-95A from the Space Telescope Science
Institute, which is operated by AURA, Inc., under NASA contract
NAS~5-26555, and by the Italian Ministry for University and Research
(MURST) under grant Cofin98-02-32.


\baselineskip=12pt

\begin{deluxetable}{lcccccccc}
\small
\tablenum{1}
\tablecaption{Physical Properties of the Jets$^{(a)}$}
\tablehead{
\colhead{Structure} & \colhead{Band} & \colhead{$m_R$} & \colhead{Dist$^{(b)}$} 
&\colhead{Width$^{(c)}$}  &\colhead{ $F_R^{(d)}$} & \colhead{ $\alpha_{RO}$} 
& \colhead{size$^{(e)}$} & \colhead{B$^{(f)}$} 
}
\startdata
\multicolumn{8}{c}{PKS 0521-365} \nl
\hline 
Jet total &R&$ 19.2\pm0.2 $&$ \dots$&$ \dots$& 106   &$ -0.73\pm 0.05 $&$ 28.6 $&$ 5.6\pm0.7$\nl 
Jet - A   & &$ 19.9\pm0.3 $&$ \dots$&$ \dots$&  46   &$ -0.67\pm 0.06 $&$ 22.8 $&$ 4.6\pm0.6$\nl
A         & &$ 19.9\pm0.2 $&$ 1.78 $&$ 1.6  $&  60   &$ -0.70\pm 0.05 $&$ 5.9  $&$ 7.5\pm1.5$\nl
B1+B2+B3  & &$ 21.7\pm0.3 $&$ 5.61 $&$ 1.3  $& (19)  &$(-0.8 \pm 0.1) $&$ 6.9  $&$ (6\pm1.5$)\nl
D         & &$ 22.0\pm0.5 $&$ 2.94 $&$ 1.6  $& (18)  &$(-0.8 \pm 0.1) $&$ 1.6  $&$ (5\pm1.5$)\nl
E         & &$ 22.3\pm0.3 $&$ 3.77 $&$ 1.4  $&  (9)  &$(-0.8 \pm 0.1) $&$ 2.2  $&$ (5\pm1.5$)\nl
\hline
\multicolumn{8}{c}{3C 371}\nl
\hline
Jet total &V&$ 20.9\pm0.2 $&$ \dots$&$ \dots$&$ 181 $&$ -0.75\pm 0.06 $&$ 7.04 $&$ 5.8\pm0.9$\nl 
A         & &$ 21.7\pm0.3 $&  3.13  &$ 1.15 $&$ 93  $&$ -0.76\pm 0.05 $&$ 2.05 $&$ 6.8\pm0.8$\nl
B         & &$ 23.1\pm0.7 $&  1.68  &$ 0.7  $&$     $&$(-0.76)        $&$ 0.44 $&$ 6.9\pm2.2$\nl
C         & &$    \dots   $&  0.46  &$ \dots$&$\dots$&$  \dots        $&$\dots $&$  \dots   $\nl
D         & &$ 24.3\pm0.8 $&  4.52  &$ \dots$&$\dots$&$  \dots        $&$\dots $&$  \dots   $\nl
\hline
\multicolumn{8}{c}{PKS 2201+044}\nl
\hline
A         &R&$ 24.2\pm0.1 $&  2.12  &$ 0.9  $&$  9  $&$  -0.85\pm0.05 $&$ 0.10 $&$ 8\pm1    $\nl
\enddata
\tablenotetext{a}{ Values depending on estimated radio fluxes are
reported in brackets.}
\tablenotetext{b}{ Distance in arcsec from center, $1\sec$ = 1.48,
1.37, and 0.74 kpc for  PKS 0521-365, 3C371, and PKS 2201+044, respectively. 
Distance for knot B of PKS 0521-365 corresponds to knot B2.}
\tablenotetext{c}{ Jet width at zero intensity in arcsec orthogonal 
to the jet axis. Due to the noise the zero intensity is $\sim 1/20$ 
of the value in the axis.}
\tablenotetext{d}{ Radio flux in mJy. For PKS 0521-365 2cm radio flux from Keel 1986. 
Fluxes for structures B, D, and E were estimated by us directly from the radio 
map of Keel (1986), and are uncertain.
For 3C371, 1.6 GHz radio flux from Akujor \ea (1994).
For PKS 2201+044, 5 GHz radio flux estimated by us from the radio map of
Laurent-Muehleisen \ea (1993).}
\tablenotetext{e}{ Volume of the structure in kpc$^3$.}
\tablenotetext{f}{ Intensity of the equipartition magnetic field in $10^{-5}$ Gauss.}
\normalsize
\end{deluxetable}

\newpage

{\bf Figure Captions}

\noindent
{\bf Fig. 1}
{\bf Upper panel:} The F702W image of the optical jet of PKS 0521-365,
obtained after subtracting the central point source and a galaxy
model. To improve the visibility of jet features, a Gaussian filter with
$\sigma = 3$ pixels was applied, reducing the resolution to 0.14
arcsec.  {\bf Lower panel:} Contour plot of the above image, with 
knots labeled.  North and east are as indicated. The big
cross at the upper right indicates the position of the central point
source. Note that the diffuse emission detected south-west of the
point source was never reported before and can actually be due to 
imperfect subtraction of the host galaxy.  Isophotes are 0.25
magnitudes apart, starting from $\mu_R = 21.9$ mag/arcsec$^2$.

{\bf Fig. 2:}
The optical image of the PKS 0521-365 jet (gray scale),
over-plotted with contours of the VLA 2 cm radio map which has
resolution of $0.\sec 3$ (Keel 1986). 

{\bf Fig. 3:}
{\bf Upper panel:} The F555W gray-scale image of the optical jet of
3C371, obtained after subtracting the central point source and a
galaxy model, smoothed with a Gaussian filter with
$\sigma = 3$ pixels, reducing the resolution down to 0.14
arcsec.  Over-plotted on the optical image is a contour plot of the
1.6GHz radio map by Akujor \ea (1994).  {\bf Lower panel:} Contour
plot of the above optical image. North and east are as indicated. The
big cross indicates the position of the central point source.
Isophotes are 0.25 magnitudes apart, starting from $\mu_V = 23.3$
mag/arcsec$^2$. 

{\bf Fig. 4:}
{\bf Upper panel:} The F702W image of the optical jet of PKS 2201+044.
In this source the central point source is not very bright, so here we
show the image after subtraction of the galaxy only. The point source
(nucleus) is characterized by the 4 diffraction spikes, not to be
confused with the jet. To improve the visibility of faint structures, 
a Gaussian filter with $\sigma = 1$ pixel was applied.  {\bf Lower Panel:}
Contour plot of the above image. North and east are as indicated. Knot
A is $2\sec .12$ from the nucleus. Two more knots most probably
associated with the jet are also marked.  Isophotes are 0.5 magnitudes
apart, starting from $\mu_R =23.2$ mag/arcsec$^2$.

{\bf Fig. 5:}
The WFPC2 F702W gray-scale image of the jet of PKS 2201+044,
over-plotted with contours of the 5GHz radio map from
Laurent-Muehleisen \ea (1993), which has lower resolution than the HST
data. The HST image has been rotated so that north is up and east on the left.

\begin{figure}
\psfig{file=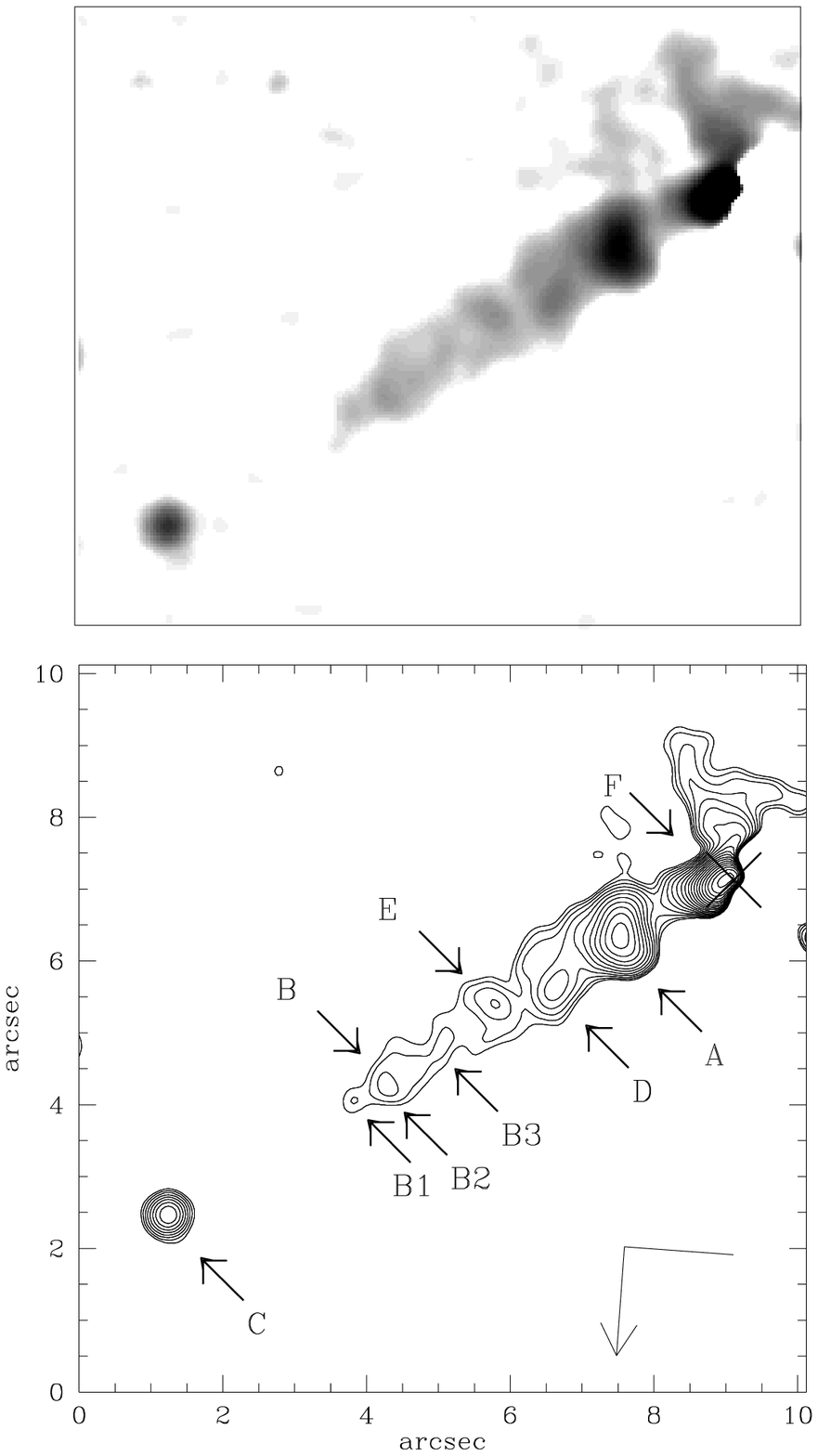,width=12cm}
\end{figure}

\begin{figure}
\psfig{file=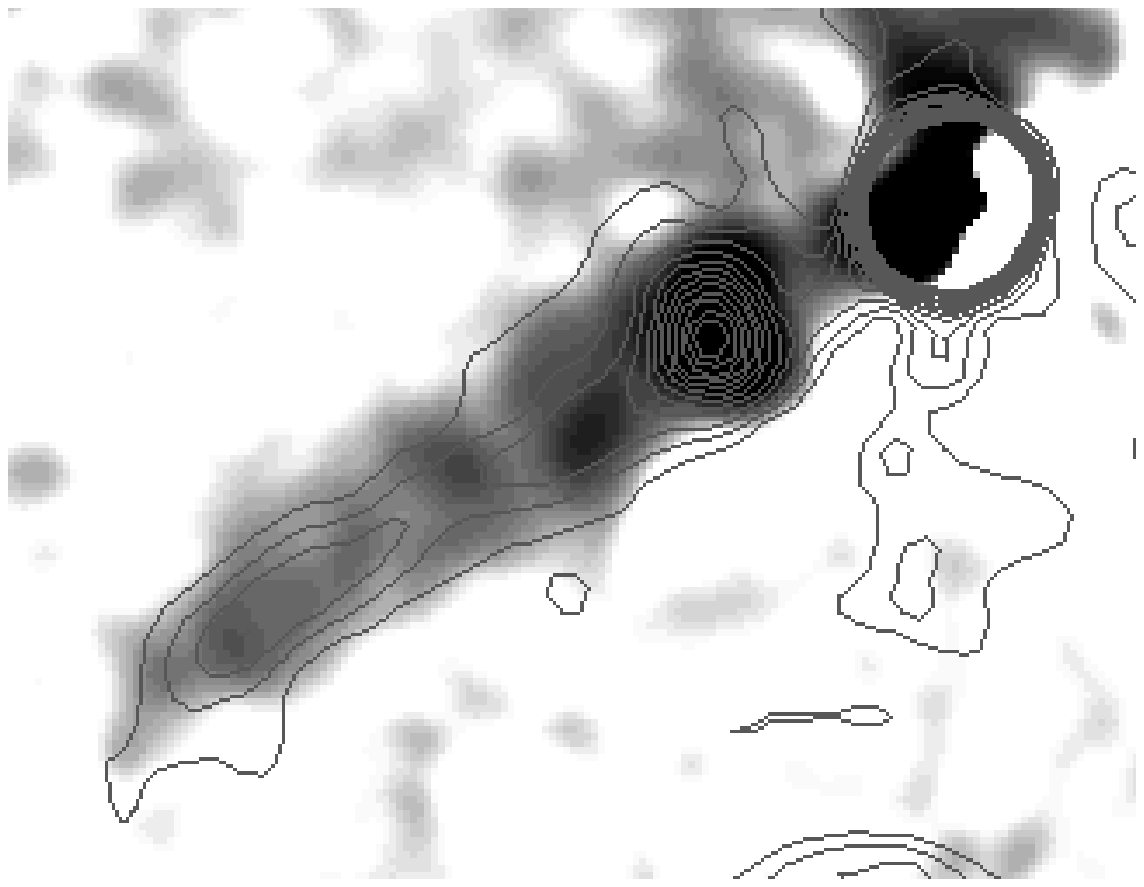,width=12cm}
\end{figure}

\begin{figure} 
\psfig{file=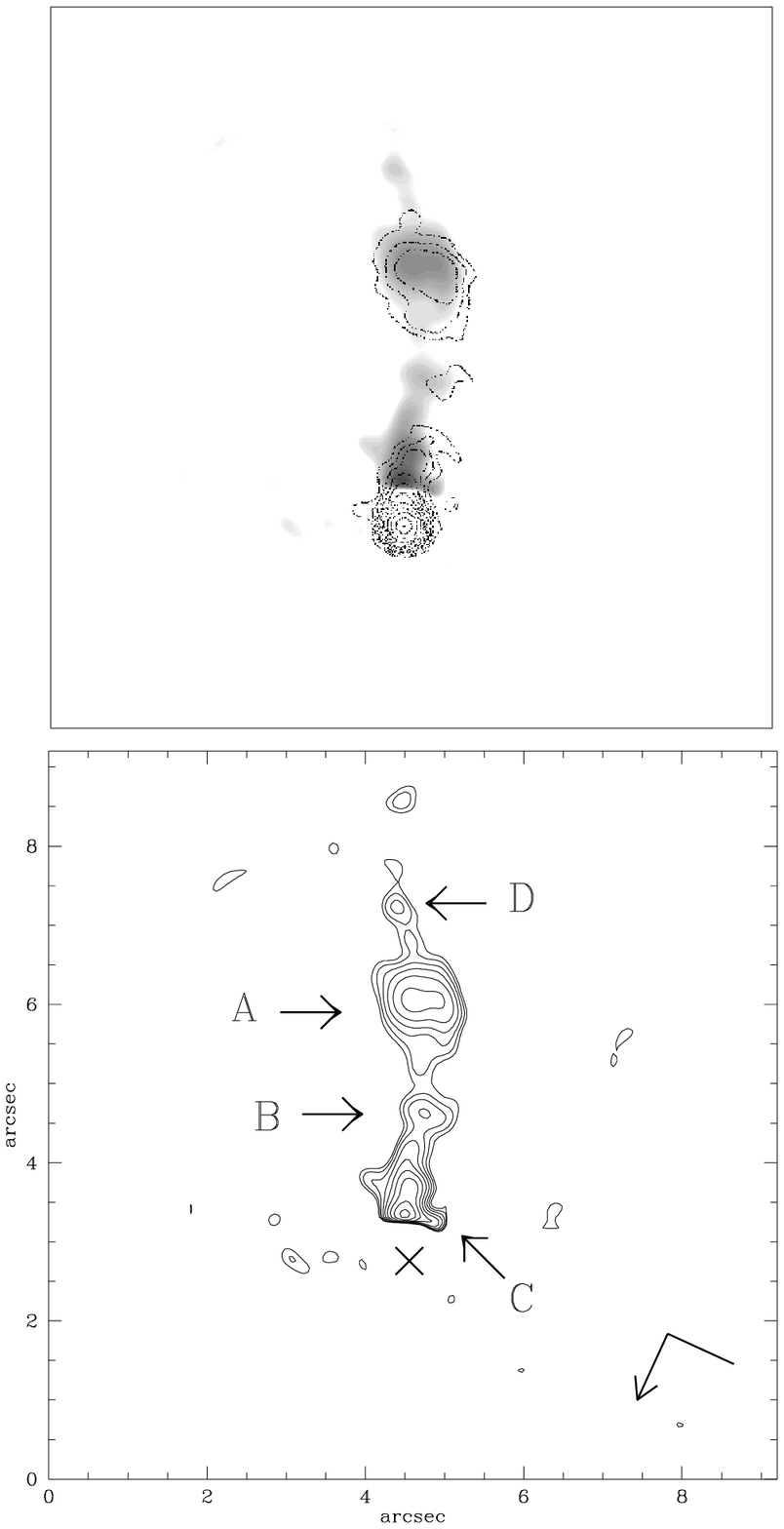,width=10cm}
\end{figure}

\begin{figure} 
\psfig{file=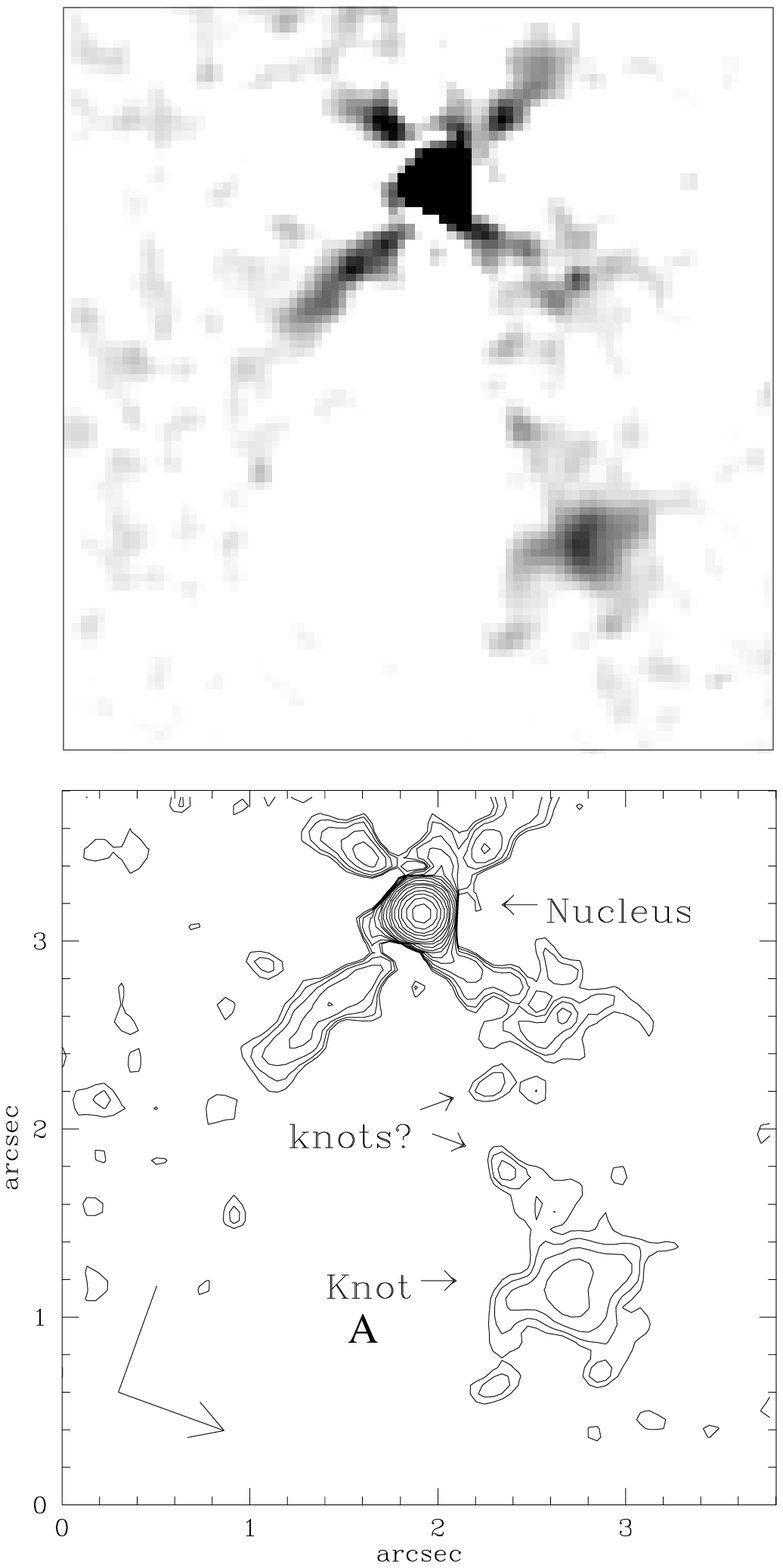,width=10cm}
\end{figure}

\begin{figure} 
\psfig{file=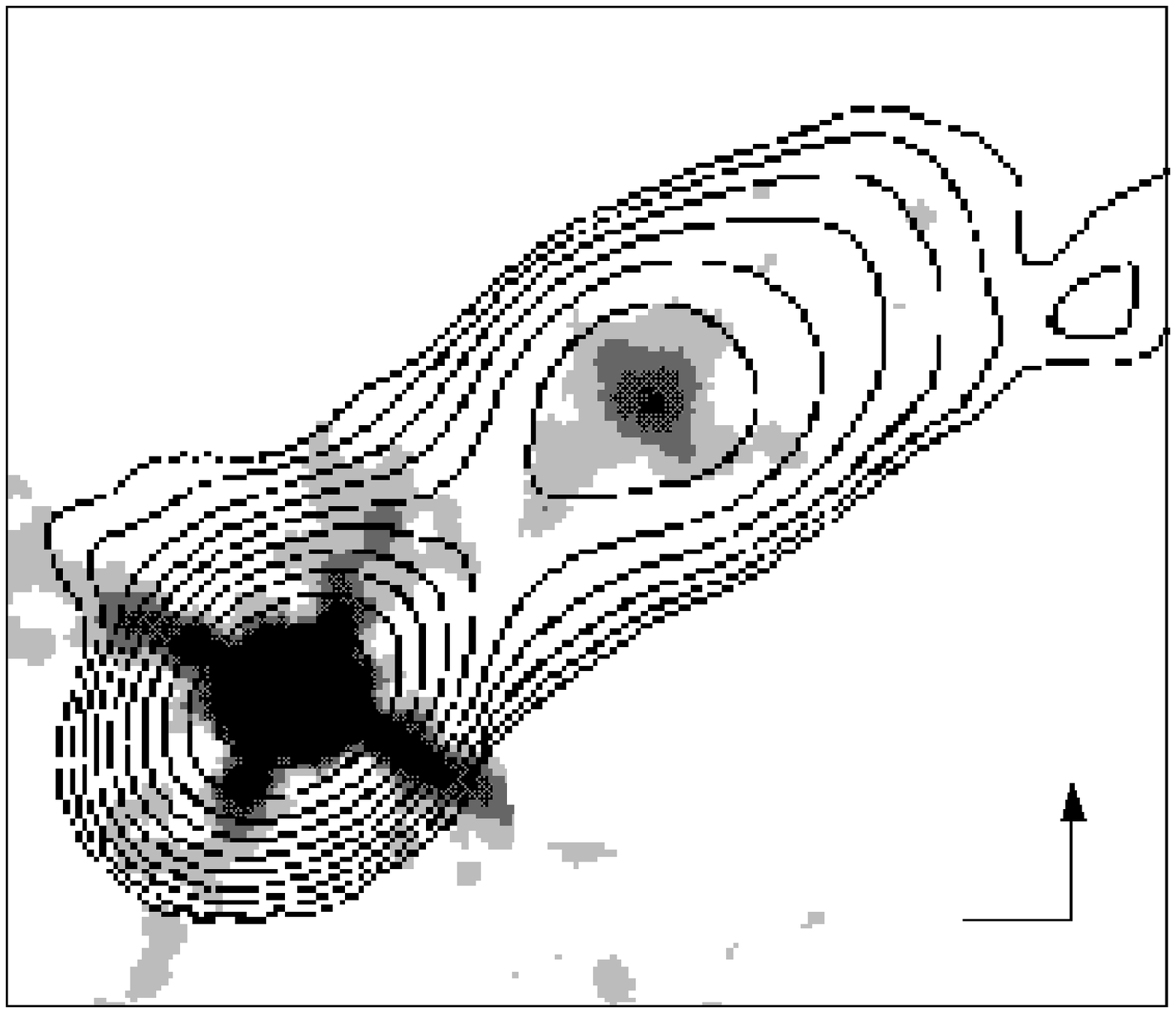,width=12cm}
\end{figure}

\end{document}